\documentclass[aps,prl,reprint,superscriptaddress]{revtex4-1}
\usepackage{graphicx}
\usepackage{amsmath}
\usepackage{amsfonts}
\usepackage{bm}
\usepackage[colorlinks=true,allcolors=blue]{hyperref}
\usepackage[usenames,dvipsnames]{color}

\usepackage{ulem}
\renewcommand{\emph}[1]{\textit{#1}}

\begin{document}

\title{No unjamming transition in a marginal vertex model of biological tissue}

\author{Daniel M. Sussman}\email{dmsussma@syr.edu}
\thanks{These two authors contributed equally to this work}
\affiliation{Department of Physics, Syracuse University, Syracuse, New York 13244, USA}
\author{Matthias Merkel}\email{mmerkel@syr.edu}
\thanks{These two authors contributed equally to this work}
\affiliation{Department of Physics, Syracuse University, Syracuse, New York 13244, USA}

\date{\today}

\begin{abstract}
Vertex models are a popular approach to modeling the mechanical and dynamical properties of dense biological tissues, describing the tissue as a network of connected polygons representing the cells. Recently a class of two-dimensional vertex models was shown to exhibit a disordered rigidity transition controlled by the preferred cellular geometry, echoing experimental findings. An attractive variant of these models uses a Voronoi tessellation to describe the cells and endows them with a non-equilibrium model of cellular motility, leading to rich, glassy behavior.  This glassy behavior was suggested to be inextricably linked to an underlying jamming transition. We test this conjecture, exploring the low-effective-temperature limit of the Voronoi model by studying cell trajectories from detailed dynamical simulations in combination with rigidity measurements of energy-minimized disordered cell configurations. We find that the zero-temperature limit of this model has no unjamming transition.
\end{abstract}

\maketitle

How does the collective behavior of biological tissues emerge from individual cellular behavior? Tissues and cellular aggregates display a rich variety of complex, non-equilibrium phenomena, and understanding the impact of microscopic cellular interactions on mechanical tissue properties will help to better understand processes ranging from multicellular development to wound healing to cancer mechanisms \cite{Friedl2009,Brugues2014,Etournay2015}. For example, recent experiments have revealed collective rigidity transitions in dense tissues intimately related to an observed change in cell-scale geometric and material parameters \cite{Angelini2011,Sadati2013,park2015unjamming,Garcia2015,Malinverno2017}, and such transitions have been interpreted through the lens of glass-like and jamming transitions \cite{Angelini2011,Sadati2013,Schoetz2013,Fredberg2014,Bi2015,park2015unjamming,Garcia2015,Bi2016,Malinverno2017}. Strikingly, although in particulate systems the athermal jamming transition is distinct from both thermal and non-equilibrium dynamical glass transitions \cite{ikeda2012unified,fily2014freezing,berthier2014nonequilibrium}, it has been suggested that these may coincide in models for dense biological tissues \cite{Bi2016,bi2016tissue}.

A successful class of models to describe tissues on the cellular scale are ``vertex'' models, which describe tissues as polygonal or polyhedral tilings of space \cite{Honda1978,Honda2001,Hufnagel2006,Farhadifar2007,Staple2010,CHASTE,Fletcher2014,alt2017vertex,Ziherl2009,Bi2014,Bi2015}.  The degrees of freedom in these models are cell vertices, allowing for complex, non-convex cell shapes. A very recent variation describes a tissue as a Voronoi tessellation of space \cite{Bi2016}, where the degrees of freedom are the cell positions (Voronoi centers). This Voronoi constraint greatly simplifies the process of handling cell dynamics, and recent experiments have shown that Voronoi tessellations approximate epithelial cell shapes reasonably well \cite{kaliman2016limits}. Although the 2D Voronoi model has enjoyed rapid adoption by many research groups \cite{bi2016tissue,su2016overcrowding,Idema2017,giavazzi2017flocking,atia2017universal,yang2017correlating,SAMOS,Li2017}, fundamental properties of this model, and even its connections with the original vertex model, are still poorly understood.

Both vertex and Voronoi models describe forces on cells by the gradient of an effective energy written in terms of cell shapes.  Here, we use a dimensionless form of a commonly used energy functional \cite{Farhadifar2007,Staple2010,Bi2014,Bi2015,Bi2016}:
\begin{equation}\label{eq:energy}
e = \sum_{i=1}^{N}{\left[k_A\left(a_i - a_0\right)^2 + \left(p_i-p_0\right)^2\right]}\text{.}
\end{equation}
This energy is a function of the area $a_i$ and perimeter $p_i$ of each cell $i$, where the unit of length is defined such that the average cell area $\langle a_i \rangle$ is one. The parameter $k_A$ controls the cell area stiffness as compared to the perimeter stiffness, and the preferred values for cell area and perimeter are $a_0$ and $p_0$, respectively.  

The 2D \emph{vertex} model exhibits a rigidity transition controlled by the preferred perimeter $p_0$, where the transition was found to be at $p_0=p_c\equiv 3.81$ \cite{Bi2015}. This transition, which was successfully compared with experimental data on asthmatic airway epithelia, was identified by studying the athermal energy landscape of the model. In contrast, a recent series of papers have studied rigidity in the 2D \emph{Voronoi} model using non-equilibrium dynamics with cellular self-propulsion \cite{Bi2016,bi2016tissue,atia2017universal,yang2017correlating,giavazzi2017flocking,SAMOS,Li2017}. This 2D self-propelled Voronoi (2D-SPV) model exhibits a glass-like dynamical transition in the limit of low motility at $p_0\approx 3.81$ \cite{Bi2016}. This surprising result was taken to imply that the athermal 2D Voronoi model also has a rigidity transition at this point \cite{bi2016tissue}, although such a transition was not directly measured. In the context of particulate systems such a coincidence between a dynamical glass and a structural jamming transition would be particularly surprising \cite{berthier2014nonequilibrium}.

In particulate models the jamming transition can be helpfully interpreted through the lens of ``Maxwell'' or constraint counting \cite{Lubensky2015}, in which one considers the balance between a model's degrees of freedom (d.o.f.) and constraints. For instance, in soft sphere systems a small change in the density can lead to large changes in the number of contacts, and the jamming transition occurs when the system is \emph{marginal}, with contacts (serving as constraints) and d.o.f.\ exactly balanced \cite{Lubensky2015,Liu2010}. In this framework one would expect clear differences between classical jamming models and the various vertex and Voronoi models. In the 2D vertex and 3D Voronoi models the number of d.o.f.\ is always larger than the number of constraints represented by Eq.~\ref{eq:energy} (or its 3D counterpart); these models are \emph{under-constrained}, and the system is rigidified by the collective onset of residual stresses  \cite{Merkel2017}.

In stark contrast, the 2D Voronoi model described by Eq. \ref{eq:energy} is \emph{always marginal}, and a constraint-counting perspective would predict that it lacks \emph{any} athermal rigidity transition. In light of this, hypothesized connections between the 2D vertex and Voronoi models, and their jamming and glass transitions, are especially surprising.

In this work we directly test whether, like the 2D vertex model, the 2D Voronoi model has an athermal unjamming transition. We first study the shear modulus of energy-minimized states and find that it is always positive, showing that all energy minima of the 2D Voronoi model are rigid. We connect this to the marginality of the model by studying the density of vibrational modes. We also perform very-low motility simulations of the 2D-SPV model and show that our results are consistent with a distinction between the dynamical glass transition and the zero-temperature behavior. We close by discussing a special $k_A=0$ limit of the model that is more closely connected to the 2D vertex and 3D Voronoi models.


\textbf{Methods:} 
We begin with 100 initial configurations for each parameter pair $(k_A,p_0)$, created by placing $N$ cells at random positions and relaxing the system with periodic boundary conditions and fixed box dimensions. Note that for fixed box dimensions, the parameter $a_0$ does not affect our results since it only renormalizes the pressure \cite{su2016overcrowding,yang2017correlating,Merkel2017}, and we set it to $a_0=1$ \footnote{Corrections to this apply in the polydisperse case.}.

\begin{figure}
\centerline{\includegraphics[width=1.0\linewidth]{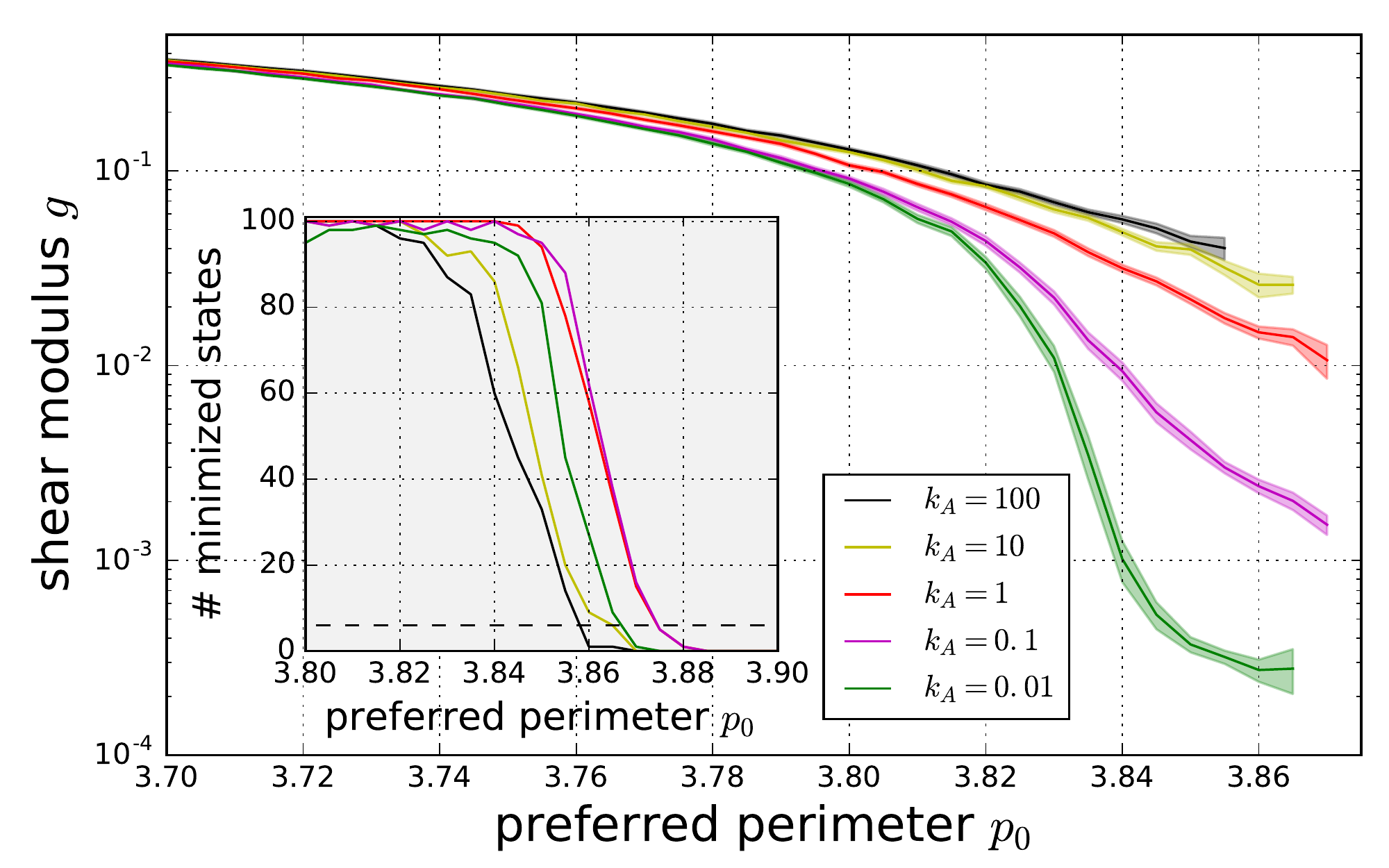}}
\caption{\label{fig:moduli} The shear modulus is strictly positive over the entire range of model parameters studied. The shear modulus $g$ is plotted versus preferred perimeter $p_0$ for increasing area moduli $k_A$ (bottom to top) in systems of $N=1024$ Voronoi cells.  The shaded regions indicate the standard error of the mean.
(inset) Number of initial states out of 100 random configurations whose energy was successfully minimized for a given $p_0$ and $k_A$. The average shear modulus was only computed if at least six minimized states were obtained (black dashed line).
}
\end{figure}

Finding disordered energy minima at large $p_0$ is highly non-trivial. We used a combination of simple gradient minimization with an adaptive step size, conjugate gradient, Newton-Raphson, and FIRE minimization \cite{bitzek2006structural}, but all of these techniques failed to consistently obtain energy-minimized states for $p_0 \gtrsim 3.87$ (Fig.~\ref{fig:moduli} inset). This failing is related to the stabilization of many-fold vertices in the ground states of the 2D Voronoi model, which leads to a very cusp-like energy landscape. Stable many-fold vertices were already observed in a thermal version of the Voronoi model \cite{su2016overcrowding}, (despite being unstable in the 2D vertex model \cite{Staple2010,Spencer2016}).  Below we restrict our numerical data to the $p_0<3.87$ regime, and comment on the high-$p_0$ regime in the discussion.

\textbf{Results:} 
In Fig.~\ref{fig:moduli} we plot the shear modulus $g$ of energy-minimized configurations as a function of $p_0$, averaged over all successfully minimized states for each parameter pair $(k_A,p_0)$. For every minimized state we observe that the shear modulus was strictly positive \nocite{dagois2012soft}\footnote{For $k_A=100$, we found 2 out of 24,200 cases where the shear modulus was below the numerical cutoff determined for the $k_A=0$ case. Studying these cases in detail, we verified that this was related to not shear-stabilizing our configurations during the energy minimization \cite{dagois2012soft}.}, directly showingthat there is no athermal unjamming transition as $p_0$ is increased. This is in contrast to the conjectured connection between rigidity transitions in the 2D Voronoi and vertex models \cite{bi2016tissue}. Despite the identical energy functional and the similarity of configurations deep in the solid phase \cite{Bi2016,kaliman2016limits}, the models thus have starkly different low-temperature properties close to the vertex model transition point $p_c$.

This lack of a mechanically rigid phase is also reflected in the eigenvalue structure of the dynamical matrix, $\mathbf{D}_{ij} = \partial^2 e/\partial \bm{r}_i \partial \bm{r}_j$, where $\bm{r}_i$ denotes the vector position of cell $i$.  For all energy-minimized states we computed the matrix elements of $\mathbf{D}$, its eigenvalues $\lambda_i$, and its eigenvectors. Most importantly, we find that the only zero energy modes of the dynamical matrix correspond to the two (trivial) translational modes.  Thus, in the athermal 2D Voronoi model there are no available modes for cell rearrangements without energetic cost. This, again, is in contrast with the vertex model, where the energy landscape becomes flat in many directions above $p_c$, corresponding to a large number of non-trivial zero modes \cite{Bi2014,Bi2015}.

\begin{figure} 
\centerline{\includegraphics[width=1.0\linewidth]{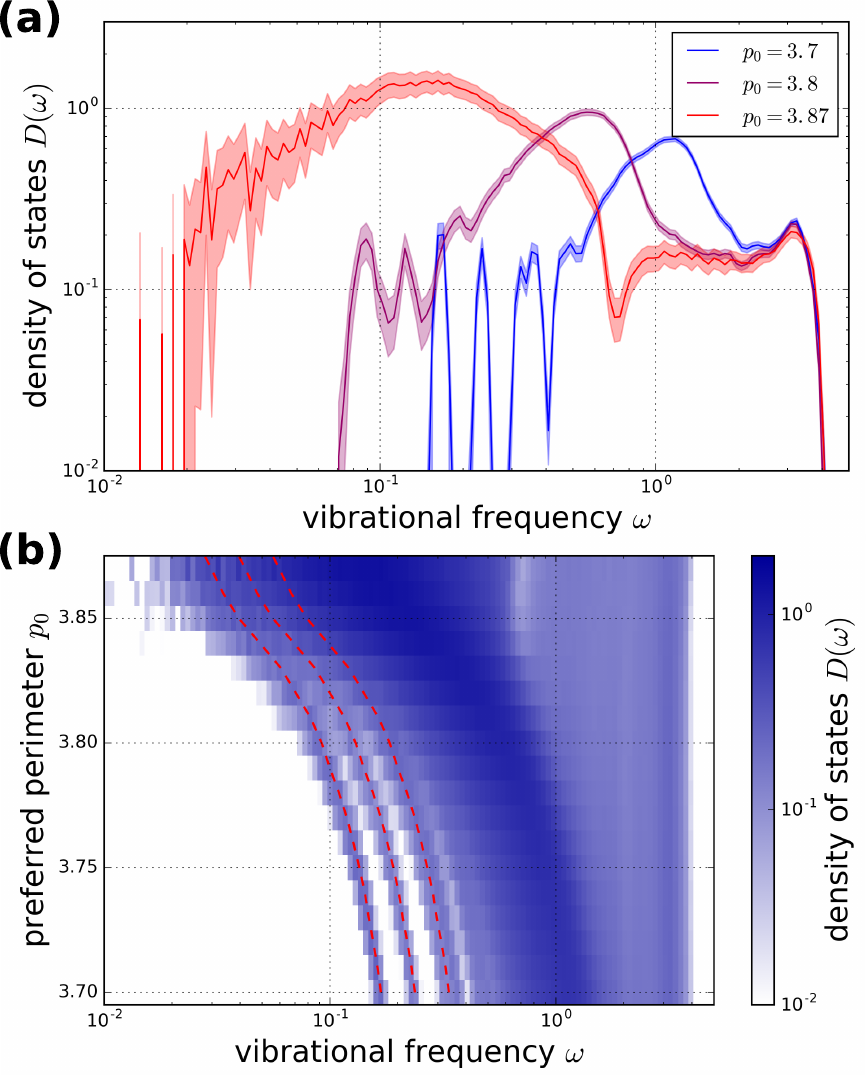}}
\caption{\label{fig:dos} The density of vibrational modes shows no signs of a sharp transition at finite $p_0$. (a) Density of states $D(\omega)$ for energy-minimized states with $k_A=1$ and system size $N=512$. From left to right the curves correspond to decreasing $p_0$. (b) Color intensity plot of $D(\omega$) for $p_0=3.7 - 3.87$.  The red dashed lines are the lowest-energy plane-wave excitations with frequencies $\omega_n=2\pi\sqrt{ng/N}$ with $n=1,2,4$, computed using the measured shear modulus $g$.}
\end{figure}

In Fig.~\ref{fig:dos}a,b we plot the density of states $D(\omega)$ of the dynamical matrix, where the frequencies are $\omega_i=\sqrt{\lambda_i}$.  Consistent with our previous findings on the shear modulus, all changes in $D(\omega)$ are smooth across the entire range of $p_0$ studied. The peaks in $D(\omega)$ at small $p_0$ are plane-wave excitations, which we verified by direct visualization, mode counting, and by computing the lowest plane-wave excitation frequencies from the measured shear modulus $g$ (red dashed lines in Fig.~\ref{fig:dos}b). At higher $p_0$ these modes hybridize with the population of disordered low-frequency modes. Here too the process is smooth, with no indication of a rigidity transition. Moreover, studying the ``unstressed network'' part of $\mathbf{D}$ \cite{Alexander1998} reveals a plateau of modes extending to very low frequencies (not plotted). We interpret this striking connection with observations in particulate models close to the jamming transition \cite{Silbert2005pre} as a consequence of these models' proximity to a marginal constraint counting point.

How should the lack of an athermal unjamming transition in the 2D Voronoi model be reconciled with the glassy dynamical transition at $p_0\approx 3.81$ observed in the low-motility limit in Ref.~\cite{Bi2016}? There the dynamical transition was quantified in terms of an effective diffusion constant, $D_\mathrm{eff} \equiv \lim_{t\rightarrow\infty} \langle \Delta r^2(t)\rangle/(4 t D_0)$ with $D_0 = v_0^2/(2D_r)$, and Ref.~\cite{Bi2016} found that $D_\mathrm{eff}$ consistently crossed a threshold value close to the same value of $p_0=3.81$. However, in light of the above results, such a glass-like transition cannot be due to an underlying jamming transition of the 2D Voronoi model. To resolve this issue, we use a recently developed GPU-based simulation package \cite{Sussman2017cellGPU,cellGPU} to repeat the measurements of the effective diffusion constant in the 2D-SPV model, but at a much greater resolution and numerical precision than was previously accessible (with orders of magnitude improvement in the total simulation length and system size).

\begin{figure}
\centerline{\includegraphics[width=1.0\linewidth]{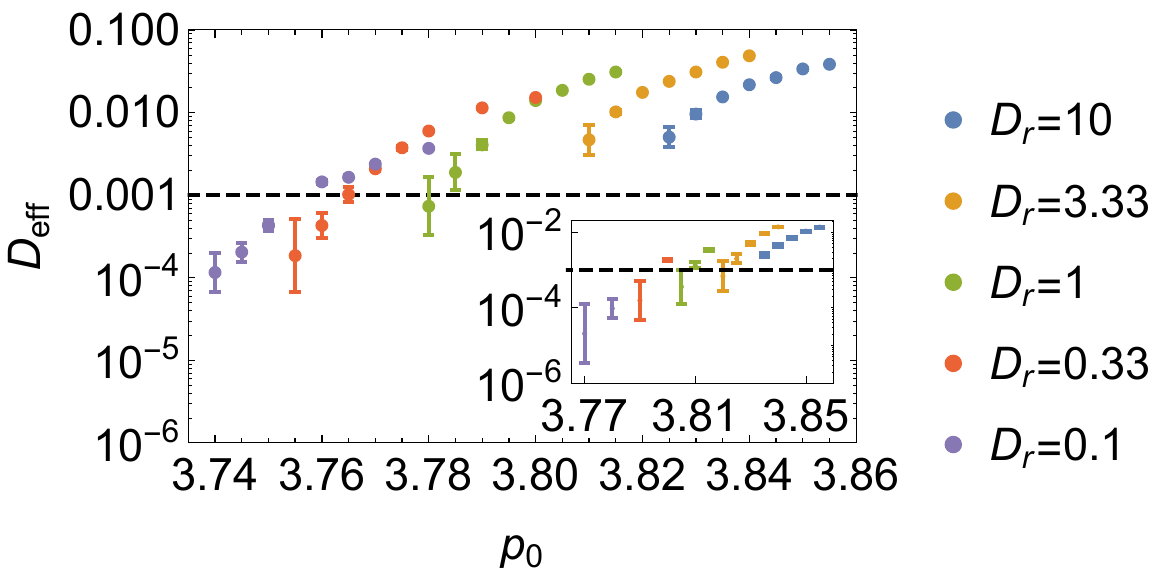}}
\caption{\label{fig:Deff} The dynamical transition point does not converge to a single value at low effective temperatures. The effective diffusion coefficient $D_\mathrm{eff}$ of the 2D-SPV model for $v_0 =0.1$ and (inset) $v_0 = 0.05$ is plotted versus $p_0$, measured at $t=5\times 10^5$ for a system size $N=5000$. The error bars are the standard error of the mean, and the dashed horizontal lines correspond to the cutoff value of $D_\mathrm{eff}=10^{-3}$ suggested in Ref.~\cite{Bi2016}.}
\end{figure}

We simulated the 2D-SPV equations of motion, $\mathrm{d}\bm{r}_i/\mathrm{d}t = \bm{f}_i + v_0\bm{n}_i$ using a simple Eulerian scheme with a time step of $\Delta t = 0.01$. The force on cell $i$ is given by $\bm{f}_i = -\nabla_i e$ with $e$ from Eq.~\ref{eq:energy}, the parameter $v_0$ is a self-propulsion speed, and $\bm{n}_i$ is a polarization vector assigned to every cell $i$ which diffuses on the unit circle with rotational diffusion constant $D_r$ \cite{Henkes2011, Bi2016}. In contrast with previous results, we find that $p_0=3.81$ no longer plays a special role, and further that the implied dynamical transition is a strong function of $D_r$ (Fig.~\ref{fig:Deff}). Hence, simulating for much longer times suggests the absence of a unique low-temperature transition point for the 2D-SPV model, thus corroborating the absence of an unjamming transition in the athermal 2D Voronoi model.

\begin{figure}
\centerline{\includegraphics[width=1.0\linewidth]{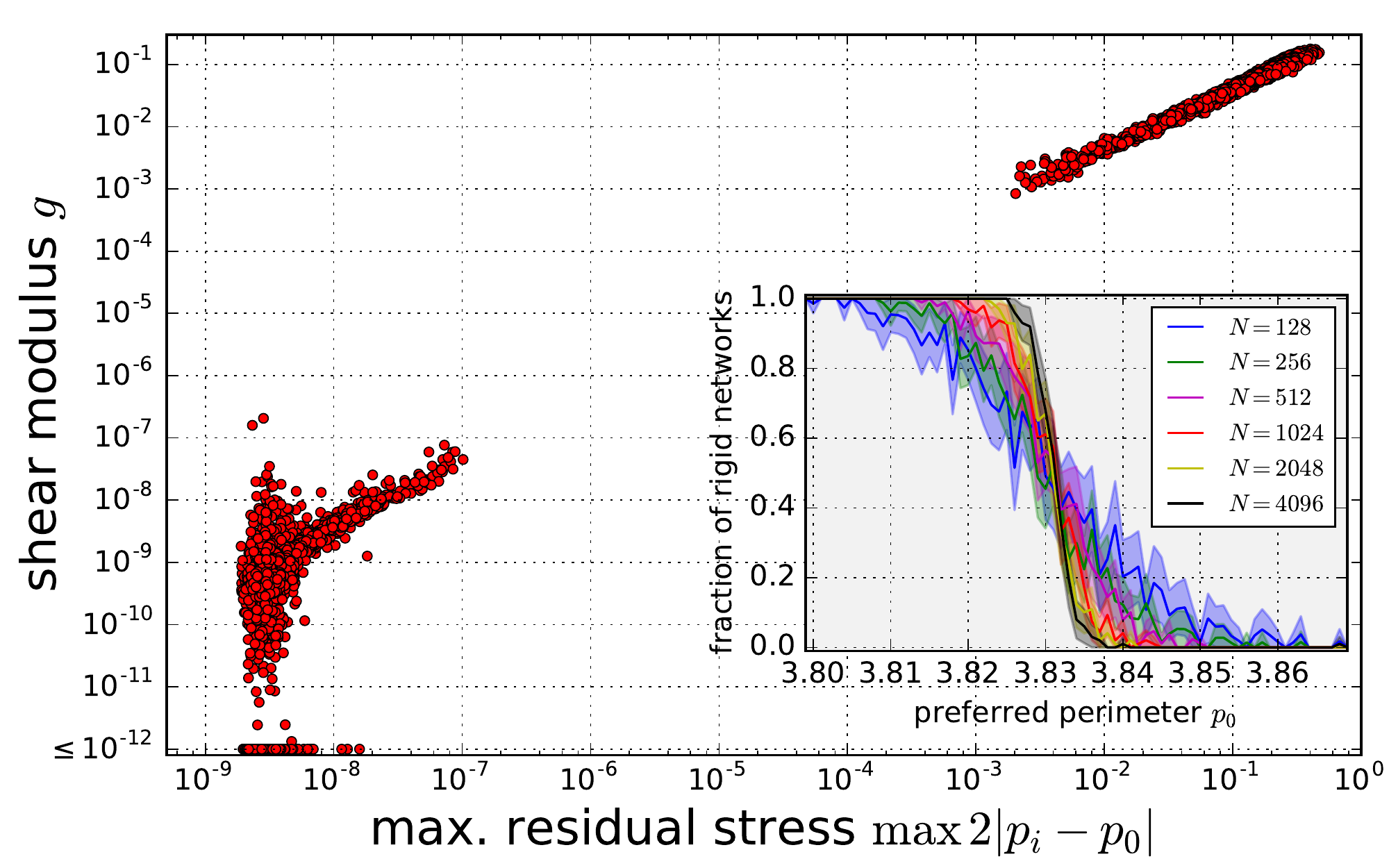}}
\caption{\label{fig:kA=0} 
The special case $k_A=0$ of the 2D Voronoi model does show a residual-stress-controlled rigidity transition. The shear modulus $g$ is plotted versus the maximal magnitude of residual stresses, given by twice the maximal deviation of cell perimeters $p_i$ from $p_0$. A clear separation between solid configurations with finite residual stresses and floppy configurations without residual stresses is visible (the inset does not cover any data points).  (inset) Quantification of the $k_A=0$ transition point based on the fraction of rigid, energy-minimized states as a function of $p_0$ for varying system size $N$.
}
\end{figure}

Finally, we comment that the limit $k_A=0$ is special and \emph{does} possess an athermal unjamming transition, which is readily explained from a constraint-counting perspective.  While the 2D Voronoi model is marginal in the general case $k_A>0$, the special case $k_A=0$ is \emph{under-constrained}, possessing $2N$ d.o.f.\ but only $N$ constraints (each perimeter term is one constraint). This puts it in a similar class as the 3D Voronoi and 2D vertex models, which are under-constrained while also possessing an athermal unjamming transition. In particular, in the 3D Voronoi model residual stresses are both necessary and sufficient to rigidify the system \cite{Merkel2017}. We find that this is also true in the $k_A=0$ case of the 2D Voronoi model: the occurrence of residual stresses is perfectly correlated with rigidity (Fig.~\ref{fig:kA=0}). Looking back at Fig. \ref{fig:moduli}, we see that for very low $k_A$ the shear modulus indeed drops by two orders of magnitude at values of $p_0$ close to the $k_A=0$ transition, but remains finite as expected from the constraint counting. Taken together, this confirms that a jamming-based constraint counting perspective helps predict the mechanical properties of both vertex and Voronoi models.

To precisely determine the value of the transition in the limiting case $k_A=0$, we plot the fraction of rigid networks in the inset to Fig.~\ref{fig:kA=0}. For these states minimized with a simple gradient descent algorithm, we found that the average transition point $p_0^\ast=3.831\pm0.001$ did not depend significantly on system size. However, when minimizing with the FIRE algorithm we obtained a significantly different transition point of $p_0^\ast \approx 3.80$ (data not shown).  Note that such a sensitivity of the transition point to the minimization protocol is consistent with a landscape that possesses a large number of very small energy barriers, emphasizing the need for high-precision numerical studies.

\textbf{Conclusion:}
We have found that the athermal 2D Voronoi model governed by Eq.~\ref{eq:energy} does not have a disordered mechanical rigidity transition, with solid behavior found throughout the entire range of $p_0$ studied. Although we were not able to obtain energy-minimized states for target perimeters $p_0\gtrsim3.87$, the $p_0$ range studied easily covers the dynamic transition point at $p_0\approx 3.81$ previously observed in the 2D SPV and vertex models \cite{Bi2014,Bi2015,Bi2016}. Note that Ref.~\cite{Bi2016} already contains hints that states in the reported fluid regime of the 2D-SPV model may be solid in the athermal limit.  For instance, for simulations performed at $p_0=3.85$ the mean-squared displacement had a notable regime of sub-diffusive behavior \cite{Bi2016}, suggesting caging with finite energy barriers to rearrangement. Taken together, this suggests that the glass-like transition reported for the 2D-SPV model does not coincide with an underlying athermal jamming transition.

This conclusion is further supported by our simulations of the 2D-SPV model. With significantly increased simulation time and system size as compared to Ref.~\cite{Bi2016}, our simulation results strongly suggest a scenario in which a dynamic glass transition is a function of the rotational diffusion; this is in contrast to previously reported results \cite{Bi2016}, but it is consistent with the decoupling of the glass and jamming transitions observed in self-propelled particle models \cite{berthier2014nonequilibrium}. 

Although we could not minimize states  for $p_0 \gtrsim 3.87$, the existence of a mechanically floppy regime in this range is theoretically disfavored. As noted above, the 2D Voronoi model is marginal, possessing as many degrees of freedom as constraints. Consistent with earlier work \cite{su2016overcrowding}, we additionally observe the stabilization of four-fold vertices in the high-$p_0$ regime, and each of these effectively adds a constraint, shifting the system \emph{further towards rigidity}. The only way this could lead to a loss of rigidity is if a number of constraints became redundant \cite{Lubensky2015}, an improbable result given the disordered geometry of the problem. Hence, we view the existence of a fluid regime at high $p_0$ in the athermal limit as very unlikely. 

In contrast, for the special case $k_A=0$, where there are extensively more degrees of freedom than constraints, we find a mechanically floppy regime with a rigidity transition close to that observed in the 2D vertex model. Furthermore, our results show that for $k_A=0$, rigidity is necessarily and sufficiently created by residual stresses, as in the 3D Voronoi model \cite{Merkel2017}. 

The fact that the transition point for the $k_A=0$ case depends sensitively on the minimization method points to the importance of the sampling of disordered states. In this paper, we started from random initial states and thus studied instantaneous quenches from infinite temperature.  Preliminary results on finite-temperature quenches showed smaller (but still non-zero) shear moduli for $k_A>0$.  Thus, while the conclusions of our work remain unaltered, it may be interesting to systematically study this dependence of rigidity on configurational sampling.

Although we found rigid ground state configurations for all target perimeters in the $k_A>0$ Voronoi model, we have not probed the energy barriers associated with either single-cell displacements or collective, low-frequency excitations. Studying these energy barriers is a natural probe of the nonlinear mechanical response of these systems, and the precise nature of the distribution of low energy barriers for the very weak solid we find in the high-$p_0$ regime will be of great importance in understanding the low-effective-temperature limit of the 2D-SPV model.  Along with a detailed study of the thermal glassy behavior, we view this as a natural target of future study. Given the propensity of the monodisperse Voronoi model to (partially) crystallize at low $p_0$, we emphasize the importance of using polydisperse samples for future studies of the glassy behavior of Voronoi-like models.

Finally, although we do not find a sharp transition in the 2D Voronoi model, it is still intriguing that both the elastic moduli at zero temperature and the finite-activity diffusion constants change by orders of magnitude in the regime close to $p_0=3.8\pm0.1$, which is suggestive of a deeper link to the transition observed in the vertex model at $p_c=3.81$. The fact that both the $k_A=0$ limit of the Voronoi model and the vertex model \emph{do} possess a transition at nearby values of $p_0$ suggests that the behavior of the Voronoi model for $k_A>0$ may be controlled by its proximity to these avoided transitions. This suggests that, in analogy with the 3D Voronoi transition \cite{Merkel2017}, a disordered geometric minimal perimeter may be relevant in thinking about all of the 2D models.  The existence of such a minimal perimeter could also provide a robust mechanism that lets the number $p_c=3.81$ appear in experiments \cite{park2015unjamming}.

\begin{acknowledgments}
We thank Lisa Manning and Cristina Marchetti for fruitful discussions and comments on this manuscript. This work was supported by NSF-POLS-1607416 (DMS), the Alfred P.\ Sloan Foundation, the Gordon and Betty Moore Foundation, and the Research Corporation for Scientific Advancement (MM); we acknowledge computing support through NSF ACI-1541396, and the Tesla K40 used for this research was donated by the NVIDIA Corporation.
\end{acknowledgments}

\bibliography{voronoiRigidity_bib}

\end{document}